\pdfoutput=1
\documentclass[preprint,12pt]{elsarticle}

\usepackage{graphics}
\usepackage{graphicx}
\usepackage{epstopdf}
\usepackage{subfigure}
\usepackage{amssymb}
\usepackage{amsmath}

\journal{Nonlinear Dynamics}

\biboptions{sort&compress}

\makeindex

\begin{document}
\begin{frontmatter}

\title{Wavelet-based discrimination of isolated singularities masquerading as multifractals in detrended fluctuation analyses}

\author[address1]{Pawe{\l} O\'swi\c ecimka\corref{cor1}}
\cortext[cor1]{Corresponding author:}
\ead{pawel.oswiecimka@ifj.edu.pl}
\author[address1,address2]{Stanis{\l}aw Dro\.zd\.z}
\author[address3]{Mattia Frasca}
\author[address4]{Robert G\c ebarowski}
\author[address5,address6]{Natsue Yoshimura}
\author[address7,address8]{Luciano Zunino}
\author[address5,address1,address9]{Ludovico Minati}

\address[address1]{Complex Systems Theory Department, Institute of Nuclear Physics Polish Academy of Sciences, ul.~Radzikowskiego~152, 31--342 Krak\'ow, Poland}
\address[address2]{Faculty of Computer Science and Telecommunications, Cracow University of Technology, ul.~Warszawska~24, 31--155 Krak\'ow, Poland}
\address[address3]{Department of Electrical Electronic and Computer Engineering (DIEEI), University of Catania, 95125 Catania, Italy }
\address[address4]{Faculty of Materials Engineering and Physics, Cracow University of Technology, 30-084 Krak\'ow, Poland}
\address[address5]{FIRST, Institute of Innovative Research, Tokyo Institute of Technology, Yokohama 226-8503, Japan}
\address[address6]{PRESTO, JST, Saitama 332-0012, Japan}
\address[address7]{Centro de Investigaciones \'Opticas (CONICET La Plata - CIC), C.C. 3, 1897 Gonnet, Argentina}
\address[address8]{Departamento de Ciencias B\' asicas, Facultad de Ingenier\'ia, Universidad Nacional de La Plata (UNLP), 1900 La Plata, Argentina}
\address[address9]{Center for Mind/Brain Science (CIMeC), University of Trento, 38123 Trento, Italy}

\begin{abstract}
The robustness of two widespread multifractal analysis methods, one based on detrended fluctuation analysis and one on wavelet leaders, is discussed in the context of time-series containing non-uniform structures with only isolated singularities. Signals generated by simulated and experimentally-realized chaos generators, together with synthetic data addressing particular aspects, are taken into consideration. The results reveal essential limitations affecting the ability of both methods to correctly infer the non-multifractal nature of signals devoid of a cascade-like hierarchy of singularities. Namely, signals harboring only isolated singularities are found to artefactually give rise to broad multifractal spectra, resembling those expected in the presence of a well-developed underlying multifractal structure. Hence, there is a real risk of incorrectly inferring multifractality due to isolated singularities. The careful consideration of local scaling properties and the distribution of H\"older exponent obtained, for example, through wavelet analysis, is indispensable for rigorously assessing the presence or absence of multifractality.
\end{abstract}

\end{frontmatter}

\section{Introduction}
\label{Intro}

The concept of multifractality, whereby not just one but an extended spectrum of exponents is required to account for the dynamics of a system, represents one of the pillars of complex signal analysis \cite{pietronero2012,kwapien2012,vicsek1992,mandelbrot1968,peng1992}. The term was coined in the context of fully-developed turbulence \cite{frisch1985} and mathematically formalized, with the multifractal or singularity spectrum as principal characteristic, by Halsey et al. in the year 1986 \cite{halsey1986}. Within the framework of multifractal formalism, a function is decomposed into subsets, which are characterized by a H\"older exponent $\alpha$ and a fractal dimension $f(\alpha)$ \cite{halsey1986}. The identified set of H\"{o}lder exponents provides explicit information about the regularity of a time-series. When only one type of regularity is present in a signal, its statistical properties are quantified by a single H\"{o}lder exponent; this results in the convergence of the multifractal spectrum to a single point. A typical example of such a monofractal process is fractional Brownian motion (fBm) with a homogenous organization of the fluctuations. On the other hand, multifractal signals, related to intermittency phenomena and correlations heterogeneity, are characterized by an extended set of H\"{o}lder exponents and developed multifractal spectrum resembling an inverted parabola \cite{drozdz2018}. A representative example is the binomial cascade generated through the iterative and multiplicative procedure. Thus, the multifractal methodology offers an opportunity to distinguish between signals characterized by the same autocorrelation function or power spectrum, but having a different underlying organization. The pervasiveness of emergent fractal and multifractal structures has made it possible to apply this methodology fruitfully across diverse areas such as physics \cite{muzy2008,subramaniam2008}, biology \cite{ivanov1999,rosas2002,makowiec2009}, physiology and neuroscience \cite{stanley1999,franca2018}, chemistry \cite{stanley1988,udovichenko2002}, economy \cite{ausloos2002,turiel2005,kutner2019,oswiecimka2005,ruan2011,grech2013,gebarowski2019}, even linguistics \cite{ausloos2012,drozdz2016} and music \cite{jafari2007,oswiecimka2011}.\par

The study of the multifractal properties of time-series can be approached in two ways. The most common approach involves estimating the spectrum through global procedures that neglect the precise location of the singularities. These procedures, as we show in our study, work well for fractal structure with a well-developed hierarchy of singularities and densely roughness of the signals. An alternative one, based on assessing singularity locally, offers the possibility to determine also the location of the H\"{o}lder exponents. This opens up the possibility of quantifying the singular behavior of the signal whenever singularities appear isolated, e.g., when analyzed processes reveal local smoothness or outliers due to measurement error. Although this approach provides more information about the fluctuation organization, it is considerably more numerically unstable, and as such, rarely applied in practical time-series analysis. However, as we clearly demonstrate in our study, this is possibly the only way of distinguishing artefactual (or apparent) from genuine multifractality. \par

Several methods have been devised for numerically estimating the multifractal spectrum of a process given a simulated or recorded signal (in the present context, equivalent to time-series) \cite{jiang2019,xiong2017,gu2010}. The prevalent algorithms encompass the multifractal detrended fluctuation analysis (MFDFA) \cite{kantelhardt2002}, the wavelet transform modulus maxima \cite{arneodo1995b}, and its development wavelet leaders (WL) \cite{jaffard1998}. Due to their numerical stability and usual accuracy, these two methods are commonly recognized as the most reliable means of estimating the multifractal spectrum \cite{oswiecimka2006,salat2017,turiel2006}. Their accuracy has been repeatedly demonstrated via synthetic time-series having known multifractal properties \cite{oswiecimka2006}. Moreover, the results obtained with these two methods can be used as a cross-test of the validity of the assessed multifractality, since MFDFA and WL are based on different numerical methods of quantifying the multifractality. The former hinged around the scaling properties of the variance, whereas the latter is grounded on the wavelet transform for decomposing the signal and characterize its self-similar properties. Thus, in our study, we applied both approaches as complementary methodologies. Albeit distinct, these methods are all based on the well-known $q$-filtering technique, which decomposes a time-series into subsets predicated on the fluctuation amplitudes. As such, they similarly provide information about the average level of fractality across all time-series segments. While their practical usefulness is beyond question, the results should be interpreted cautiously due to the inherent complexity of both the signals under analysis and of the algorithms themselves. Here, several compelling examples are given of how even elementary systems can yield signals for which a naive interpretation of the multifractal spectrum, obtained via both the MFDFA and the WL, leads to completely flawed conclusions. To this end, consideration is given to simulated and experimentally-recorded time-series from the Saito chaos generator, a simple four-dimensional non-linear dynamical system with a strong hysteretic component, as well as to synthetic signals which by construction cannot possess any multifractal structure. The results show that relying only on the multifractal spectrum width as an indicator of multifractality can be profoundly misleading. At a minimum, other spectral attributes such as the asymmetry have to be taken into account, where the key is considering the local scaling properties and the distribution of H\"older exponents.\par

This paper is organized as follows. In Sec. II, the notion of local and global H\"older exponents is introduced, together with the methods of MFDFA and WL. In Sec. III, a selection of examples of true multifractality is firstly introduced. Then, the case of a hysteretic oscillator is considered to exemplify the consequences of isolated singularities on both MFDFA and WL, followed by the presentation of paradigmatic synthetic signals leading to similar issues. Finally, in Sec. IV, practical recommendations for the proper application of these analyses are offered.

\section{Methodology}
\label{Methods}  
\subsection{H\"older exponents} 
Analyzing the regularity of a signal provides essential insight into its statistical properties and possible underlying geometrical structure. To characterize the local singular properties of a time-series, then, the point-wise H\"older exponent $\alpha$ can be considered. Given a function $f$, for each $x_0 \in \Re$ the same is defined as follows \cite{jaffard2006a}:
\begin{equation}
\alpha(x_0)=\sup\{h: f \in C^h(x_0)\}\textrm{ ,}
\end{equation}  
where $f$ belongs to the H\"older space $C^h(x_0)$ if and only if
\begin{equation}
|x-x_0| \le \epsilon, \quad |f(x)-f(x_0)| \le C|x-x_0|^h\textrm{ .}
\end{equation}
Values of the H\"older exponent approaching zero indicate increasing irregularity of the function $f$; conversely, larger values of $\alpha$ denote more regular fluctuations. The multifractal formalism is a statistical description of functions through quantifying their distribution of the point-wise H\"older exponents, which is naturally extended to continuous signals and discretized time-series.

\subsection{Multifractal detrended fluctuation analysis} 
Multifractal detrended fluctuation analysis (MFDFA) \cite{kantelhardt2002} is a method for detecting and quantifying the scaling properties of time-series which is widely applied across diverse areas of experimental and computational science \cite{wang2014,oswiecimka2018,biswas2012,dutta2013,thompson2016,kwapien2005}. It comprises multiple steps, which may be summarized as follows. Let us consider a time-series $x_i$ having length $N$, $i=1,2...N$ and, as a first step, calculate its profile according to 
\begin{equation}
X\left(j\right) =\sum_{i=1}^j[x_{i}-\langle x \rangle]\textrm{ ,}
\label{profiles_DFA}
\end{equation}
where $\langle x \rangle$ denotes the mean of time-series $x_i$. Since fractality manifests as patterns which are self-similar across different temporal scales, the profile has to be analyzed over segments of different length. Thus, the time-series is next subdivided into $N_s$ non-overlapping segments $\nu$ of length $s$ ($N_s=\textrm{int}(N/s)$) starting from the beginning. However, since the length is not necessarily an integer multiple of the scale $s$, the procedure is also repeated starting from the end, yielding a total of $2N_s$ segments. To remove possible trends in the time-series, which can distort the results, in each segment $\nu$ a polynomial of order $m$ ($P^{(m)}_{\nu}$) is fit and subsequently subtracted from the data. The effectiveness of this detrending step strongly depends on the polynomial order, and it is generally agreed that small values of $m$ provide the most reliable results \cite{oswiecimka2013}. Here, no statistically discernible differences were found between results obtained for $m=2,3\textrm{ and }4$, thus, for brevity, only results assuming $m=4$ are given.\par
As a next step, the detrended variance is calculated within each segment according to
\begin{equation}
F^{2}(\nu,s)=\frac{1}{s}\sum_{k=1}^{s}(X((\nu-1)s+k)-P^{(m)}_{\nu}(k))\textrm{ .}
\label{detrended_variance}
\end{equation}
Then, in order to quantify the fractal properties of the signal with respect to the amplitude, the $q$-order filtering technique is applied, obtaining the $q$-order fluctuations function
\begin{equation}
F_q(s)=\Big\{\frac{1}{2N_s} \sum_{\nu=1}^{2N_s}[F^{2}(\nu,s)]^{q/2}\Big\}^{1/q}, q \in \Re \setminus \{0\}\textrm{ ,}
\label{Fluctuations_funtion}
\end{equation}
where $q$ operates as a filter which discriminates fluctuations based on their amplitude; more precisely, negative and positive settings of $q$ respectively emphasize small and large changes. Fractality in a time-series manifests itself as power-law behavior of $F_q(s)$ over different scales, that is,
\begin{equation}
F_q(s)\sim s^{h(q)}\textrm{ ,}
\end{equation}
where $h(q)$ denotes the generalized Hurst exponent. Hence, $h(q)$ represents the fractality of the fluctuations selected by a given setting of $q$. For monofractal time-series, $h(q)$ is constant and equals the Hurst exponent $h(q)=H$ \cite{hurst1951,feder1988}. This can be used to classify time-series with respect to linear correlations. Namely, $H>0.5$ indicates persistent dynamics (i.e., positive long-range correlation), whereas for $H<0.5$ a signal is anti-persistent (i.e., a tendency to reverse is observed); on the other hand, $H=0.5$ denotes the absence of any linear correlation. For multifractal signals, $h(q)$ is a decreasing function of $q$, and the Hurst exponent is retrieved at $h(q=2)=H$. Thus, for better visualizing the results and interpreting the spectrum of the generalized Hurst exponents, the same can be converted into the multifractal spectrum via the Legendre transform of the scaling function $\tau(q)=qh(q)-1$, or directly through
\begin{equation}
\alpha=h(q)+qh^{'}(q)\textrm{ ,} \quad
f(\alpha)=q[\alpha-h(q)]+1\textrm{ ,}
\label{spectrum-MFDFA}
\end{equation}
where $\alpha$ is the H\"older exponent, and $f(\alpha)$ refers to the fractal dimension of the data supported by a particular $\alpha$.\par
The intensity of multifractality, and thus the degree of signal complexity, is often quantified through the width of the multifractal spectrum, that is, $\Delta \alpha=\alpha_\textrm{max}-\alpha_\textrm{min}$. The larger $\Delta \alpha$, the more developed a multifractal structure is deemed to be. Another important feature of the multifractal spectrum is its asymmetry. For the paradigmatic case of the binomial cascade, a mathematical multifractal, $f(\alpha)$ resembles a symmetric inverted parabola \cite{calvet1997}. However, for real-world time-series, the spectrum is often asymmetric, having one side better developed than the other; this stems from a heterogeneous organization of the signal fluctuations across scales. Hence, through quantifying the spectral asymmetry, one can retrieve critical information about the temporal organization of a time-series. The asymmetry parameter is defined as \cite{drozdz2015}
\begin{equation}
A_\alpha=(\Delta\alpha_\textrm{L}-\Delta\alpha_\textrm{R})/(\Delta\alpha_\textrm{L}+\Delta\alpha_\textrm{R})\textrm{  ,}
\end{equation}
where $\Delta\alpha_\textrm{L}$ and $\Delta\alpha_\textrm{R}$ stand, respectively, for the distances between the spectral maximum and the smallest and largest values of $\alpha$. In turn, the degree of the asymmetry is quantified as $|A_\alpha|$, whereas the sign indicates the asymmetry direction. A positive value of $A_\alpha$ hallmarks a leftwards-stretched spectrum and denotes a well-developed fractal organization of the large fluctuations, while smaller ones are governed by simpler dynamics. Contrariwise, for negative $A_\alpha$ the spectrum is stretched towards the right, denoting more complex behavior of the small fluctuations compared to the larger ones.

\subsection{Multifractal analysis based on wavelet leaders}
Another class of techniques for estimating the multifractal characteristics of a non-stationary time-series is based on the wavelet transform \cite{muzy1994,jaffard2004}. According to these techniques, a signal is decomposed into the elementary space-scale wavelet coefficients by means of a family of functions stemming from a basic function, the so-called mother wavelet. By scaling and translation of the mother wavelet $\psi_{a,s}(x)=s^{-1/2}\psi(\frac{x-a}{s})$, one can obtain a decomposition of the signal at each scale $s$ corresponding to a frequency band, separately for all time-points $a$ ($a,s\in \Re, s>0$). The wavelet transform of a function $f(x)$ is defined as \cite{arneodo1995a}
\begin{equation}
Wf(a,s)=\frac{1}{s^{-1/2}}\int_{-\infty}^{+\infty}f(x)\psi\bigg( \frac{x-a}{s}\bigg)dx\textrm{  .}
\end{equation} 
Importantly, visualization of the resulting wavelet spectrum $Wf(a,s)$ on the scale-time plane promptly reveals the skeleton of the hierarchical structure of the process being analyzed. The choice of the mother wavelet is dictated by it being well-localized in both the time and frequency domains (derivatives of a Gaussian function are often used as mother wavelets). A crucial property of the wavelet transform is its close relation with the H\"older exponent $\alpha$ \cite{muzy1994}, wherein
\begin{equation}
Wf(x_0,s) \sim s^{\alpha(x_0)}, \quad s\rightarrow 0^+\textrm{  .}
\label{wavelet_scaling}
\end{equation}
Hence, a local singularity $\alpha(x_0)$ can be characterized by the scaling behavior of the wavelet transform around the point $x_0$. Moreover, the maxima of the wavelet transform produce maxima lines in space-scale half-plane, which converge towards loci of singularity. Thus, by retrieving the power-law behavior of the wavelet transform coefficients along these lines, one can estimate the H\"{o}lder exponents, and in turn, quantify the singularity strength \cite{arneodo1995b}. Due to the instability of the canonical wavelet-based multifractal methods whenever a large number of coefficients are close to zero, and due to its insensitivity to oscillating singularities \cite{lashermes2005}, the notion of wavelet leaders (WL) was introduced \cite{jaffard1998,jaffard2004}. For a discrete scale parameter $s_j=2^{-j}$ and time $a_{j,k}=2^{-jk}$ ($j,k \in Z$), the signal can be recovered via the formula \cite{wendt2009,deliege2014}
\begin{equation}
f(x)=\sum_{j,k\in Z}c_{j,k}\psi(2^{-j}x-k)\textrm{  ,}
\end{equation}
where the wavelet coefficient $c_{j,k}$ is given by
\begin{equation}
c_{j,k}=2^{-j}\int_{\Re}f(x)\psi(2^{-j}x-k)dx\textrm{  .}
\end{equation}
In this study, the Daubechies wavelet with 4 vanishing moments was used \cite{daubechies1988}. The wavelet leader of $x_0$ at the level $j$ denotes the largest wavelet coefficient among those existing in the spatial neighborhood of $x_0$ at finer scales \cite{figliola2012}. Formally, for the dyadic interval $\lambda_{j,k}=[2^jk,2^j(k+1)]$, it is defined as
\begin{equation}
L_j(x_0)=\sup_{\lambda' \subset 3\lambda_{j,k}(x_0)}|c_{j,k}(\lambda')|\textrm{  ,}
\end{equation}
where $3\lambda_{j,k}(x_0) = \lambda_{j,k-1}\cup\lambda_{j,k}\cup \lambda_{j,k+1} = [2^j(k-1,2^j(k+2)))$ and contains $x_0$. For a given scale $2^j$, one can define structure functions $S(q,j)$ based on the $q$-th order average of the leaders
\begin{equation}
S(q,j)=2^{j}\sum_{\lambda \in \Lambda_{j}}L_{j}^q\textrm{   ,}
\label{partition_function}
\end{equation}
where $q$ is a real number, and $\Lambda_j$ is a set of dyadic intervals at scale $j$. Power-law behavior of the structure function in the limit of small scales $S(q,j)\approx C_q2^{j \zeta (q)}$, ($2^j\rightarrow0$) is a manifestation of scale invariance. Thus, $\zeta(q)$ determines the scaling exponents, and can be numerically estimated by means of a log-log regression. Since the $\zeta(q)$ function is necessarily concave \cite{jaffard2006}, the Legendre transform can be used to estimate the multifractal spectrum according to the formula
\begin{equation}
f(\alpha)=\inf_{q\in \Re}(q\alpha-\zeta(q))+1\textrm{   .}
\label{spectrum-LW}
\end{equation}

\subsection{Global vs. local H\"older exponents}
The multifractality of a time-series manifests itself through sets of non-trivial H\"older exponents, which quantify the local variation in its irregularity \cite{ayache2004}. These exponents may be collectively quantified by means of a ``global'' measure, obtained from the multifractal spectrum in (Eq.~(\ref{spectrum-MFDFA}) or (Eq.~\ref{spectrum-LW}) and denoted as $\alpha_\textrm{G}$, or directly through the analysis of local scaling properties of the signal by means of Eq. (\ref{wavelet_scaling}) and denoted as $\alpha_\textrm{L}$. Ideally, the two approaches should give consistent results. However, as demonstrated below, the multifractal analysis of complex time-series has limitations which, under certain circumstances, yield misleading signatures of multifractality.\par

\subsection{Surrogates}
To assess the statistical validity of the results, additional analyses were performed on surrogate time-series for all scenarios under consideration. Two commonly-used surrogate sets were generated. One relies on randomization of the Fourier phases and, as such, preserves only the spectral amplitudes while obliterating all non-linear inter-dependencies \cite{schreiber2000}. The other involves randomly permuting the time-points, destroying all temporal correlations while preserving the value distribution \cite{oswiecimka2005}. 

\section{Results}

\subsection{Examples of truly multifractal time-series}
\begin{figure}[h!]
\centering
\includegraphics[scale=0.4,keepaspectratio=true, trim={1.6cm 0 0cm 0}]{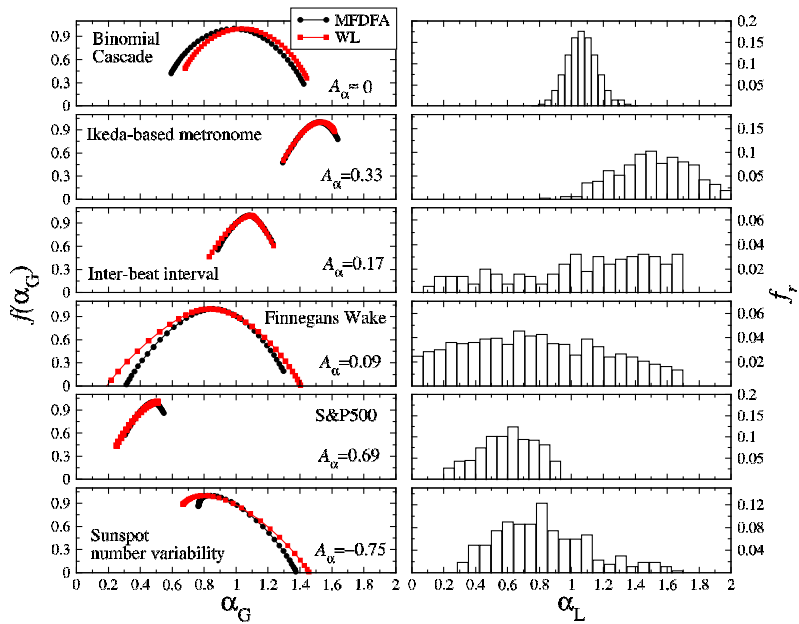}
\caption{Examples of true multifractality. (a) Multifractal spectra and (b) relative frequency $f_r$ histograms of the H\"older exponents. The following cases are presented: binomial cascade, chaotic metronome derived from the Ikeda map, inter-beat intervals, sentence length in ``Finnegans Wake'', logarithmic returns of the S\&P500 index, and sunspot number variability.}
\label{Fig1}
\end{figure}

For comparison with the particular cases considered below, representative instances of real multifractals having diverse properties are firstly presented, based on analyses conducted over the range $q\in[-4,4]$ \cite{drozdz2009}. To this end, two mathematical multifractals are considered, namely the binomial cascade and the chaotic metronome derived from the Ikeda map \cite{stephen2011}, together with several real-world time-series: the inter-beat intervals extracted from electrocardiographic signals (103885 data points), the sentence length variability of the ``Finnegans Wake'' book by James Joyce, the logarithmic returns of the American stock market index S\&P500 (7440 data points), and the sunspot number variability (43495 data points) \cite{oswiecimka2006,ivanov1999,drozdz2016,oswiecimka2005,drozdz2015}. In all these cases, the multifractal spectrum $f(\alpha_\textrm{G})$ assumes the shape of a wide inverted parabola, spanning $\Delta \alpha_\textrm{G}>0.2$, indicating a multifractal organization of the data (Fig.~\ref{Fig1}, left). Yet, the spectra develop different degrees of asymmetry. For the binomial cascade, the inter-beat intervals, and the sentence length variability, the spectra appear almost symmetrical ($A_\alpha \approx 0$), which suggests a homogeneous distribution of the correlations over small and large fluctuations. On the other hand, for stock market data and sunspot number variability, the asymmetry is, respectively, positively- and negatively-skewed. Thus, multifractality of the S\&P500 price variation is mainly the effect of a complex organization of the large fluctuations, whereas the arrangement of small fluctuations is primarily responsible for multifractality in the time-series of sunspot numbers.\par
Importantly, the presence of true multifractality is confirmed, for all these cases, via analysis of the local scaling properties (Fig.~\ref{Fig1}, right). Therein, a continuous distribution of the estimated H\"older exponents spans a range of $\alpha_\textrm{L}$ even broader than in the multifractal spectrum, incidentally revealing the higher sensitivity of the wavelet transform on the local scaling properties compared to the global methodology, which mainly reflects the prevalent singularities in the time-series.   
 
\subsection{Artefactual multifractality in the Saito chaos generator}

To illustrate the potential pitfalls inherent in drawing hasty conclusions solely from global measures, let us now consider the case of the Saito chaos generator, which is a four-dimensional non-linear oscillator consisting of the following dimensionless state equations \cite{saito1990}:
\begin{equation}
\label{Saito_eqx}
\begin{cases}
 \dot{x}=-z-w\\
 \dot{y}=\gamma(2\delta y+z)\\
 \dot{z}=\rho(x-y)\textrm{  ,}\\
 \dot{w}=(x-h(w))/\varepsilon
 \end{cases}
\end{equation}
wherein
\begin{equation}
 h(w)= 
 \begin{cases}  
     w-(1+\eta) & \text{if } w \ge \eta \\
     -\eta^{-1}w       & \text{if } |w| < \eta \quad \\ 
     w+(1+\eta)  & \text{if } w \le -\eta\textrm{  .}
     \label{Saito_eqw}
  \end{cases}
\end{equation}
Despite its simple form and low dimensionality, this system readily generates rich dynamics spanning periodicity, quasi-periodicity, chaos, and eventually hyperchaos as a function of the parameters $\gamma$, $\delta$, $\varepsilon$, $\eta$, and $\rho$. Here, it was initially deemed of interest from the perspective of its hypothetical ability to generate signals having a truly multifractal structure; however, in the course of numerical investigation, another feature was realized to be fundamentally important for the purposes of the present work, namely, the presence of the hysteresis function $h(w)$, which only enters the equation of the state variable $w$. As a consequence of it, even though all state variables conjointly participate in the temporal dynamics, $x,\ y,\ z$ have rather smooth an activity, whereas, in the limit of $\varepsilon\rightarrow0$, the temporal evolution $w$ is characterized by sudden jumps. As shall become clear, it is these discontinuities, namely the combination of slow and fast motions corresponding to the continuous manifold and sudden jumps, which may lead to a mistaken inference of multifractality. Unless indicated otherwise, the parameters were set for operation in the hyperchaotic regime, that is, $\gamma=1$, $\delta=0.94$, $\varepsilon=0.01$, $\eta=1$, and $\rho=14$ \cite{saito1990,buscarino2017}.\par

Preliminary examination revealed differences between short- and long-range temporal correlations in the simulated time-series, giving rise to a cross-over in the fluctuation functions. In particular, the multiscale characteristics revealed a strong autocorrelation only over short time scales (i.e., $s<1000$), occurring alongside a monofractal organization with weak linear correlations on the larger scales. Thus, in the analyses below, the focus is on the short-range correlations, which are relevant to the search for possible multifractality.

\subsubsection{Numerical simulations}

\begin{figure}[h!]
\centering
\includegraphics[scale=0.35,keepaspectratio=true, trim={1cm 0 0 1cm}]{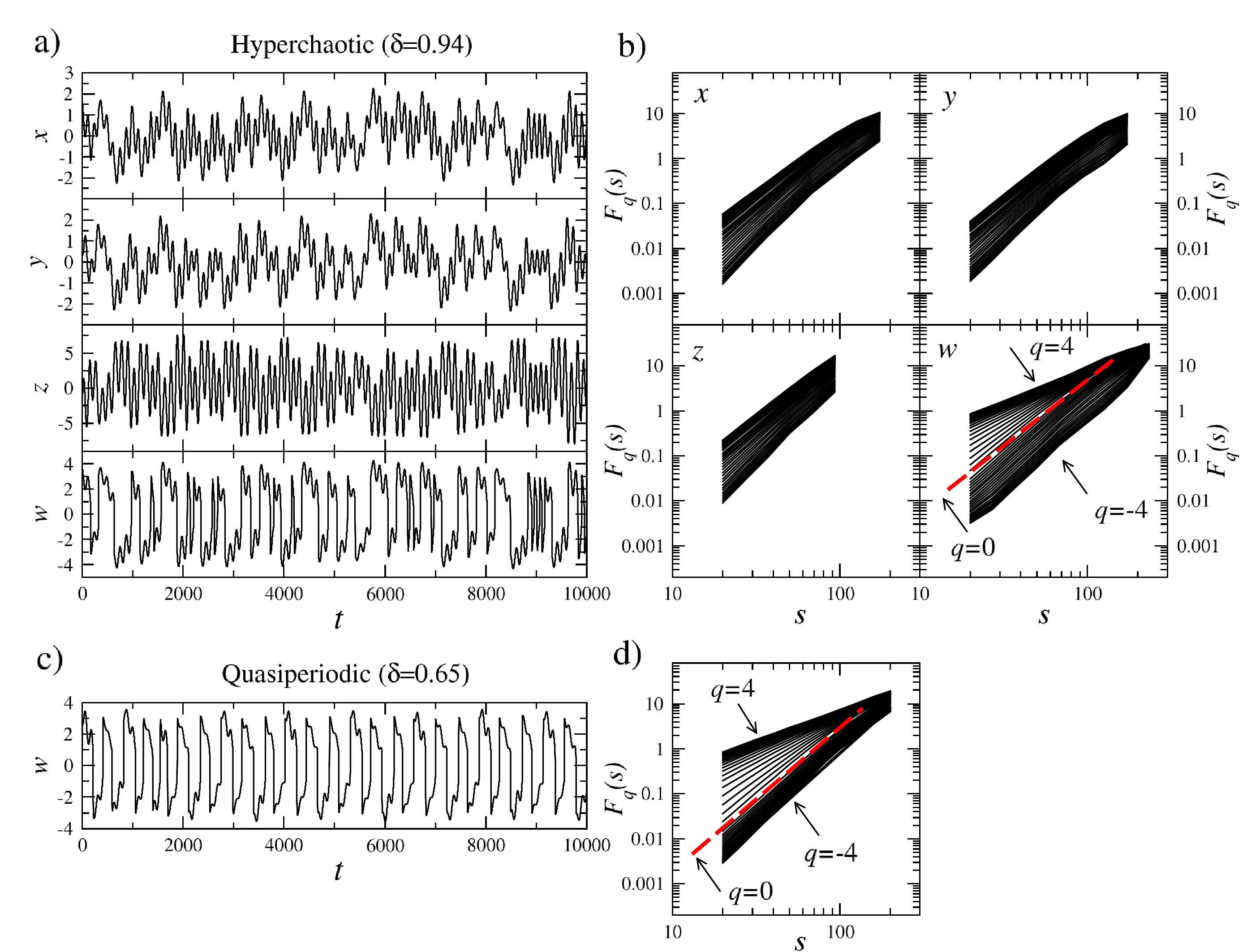}
\caption{Dynamics of the simulated Saito chaos generator. (a) Time-series in the hyperchaotic regime (all variables), and (b) corresponding fluctuation functions $F_q(s)$. (c) Time-series in the quasiperiodic regime ($w$ only), and (d) corresponding fluctuation functions.}
\label{Fig2}
\end{figure}

Time-series having a length of $10^6$ points were simulated given equations (\ref{Saito_eqx})-(\ref{Saito_eqw}), applying the adaptive step-size Runge-Kutta (4,5) method and returnig the results at a fixed step size of $\Delta t=0.1$ \cite{dormand1980}. All simulations were repeated 10 times with randomized initial conditions. Representative segments for each variable in the hyperchaotic regime are depicted in Fig. \ref{Fig2}a. Evidently, the dynamics of $x,\ y,\ z$ are characterized by the markedly irregular behavior characteristic of chaotic systems. However, the dynamics of $w$ are even more complex, featuring sharp upward and downward jumps. Even though the underlying system is the same, the multifractal properties of the signals, being influenced by the presence of singularities, could then be partially dependent on the variable under consideration. This observation is confirmed by the corresponding fluctuation functions $F_q(s)$ (Fig.~\ref{Fig2}b). Therein, it is clearly visible that the functions obey power-law behavior, which is a signature of fractal organization: however, while for $x,y$ and $z$ the scaling is rather homogeneous, for $w$ a pronounced heterogeneity is apparent. Moreover, for the latter the majority of fluctuation functions have a slope close to those found close for the extreme values of $q$, i.e., $q=\pm4$; only a minority assume intermediate levels, a fact that already points to a more bifractal-like organization of the data rather than to a well-developed multifractal structure.\par

Strikingly, rather similar characteristics of $F_q(s)$ are observed in the quasiperiodic regime, with $\delta=0.65$ (Fig. \ref{Fig2}c; for brevity, results are only shown for $w$). Though the dynamics are profoundly different compared to the hyperchaotic regime, the heterogeneity of the fluctuations functions remains most pronounced for $w$, with the distribution of slopes nearly unchanged and characteristic of a bifractal structure (Fig.~ \ref{Fig2}d).

\begin{figure}[h!]
\centering
\includegraphics[scale=0.35,keepaspectratio=true,trim={1.5cm 0 0 0}]{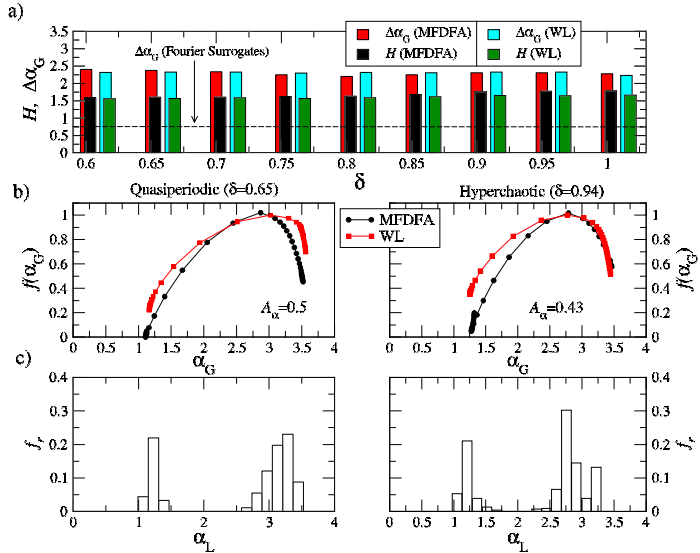}
\caption{Analysis of the simulated $w$ time-series from the Saito chaos generator. (a) Width of the multifractal spectrum $\Delta \alpha$ and Hurst exponent given different settings of the control parameter $\delta$. Horizontal dashed line: average spectra widths estimated for Fourier-based surrogates; for randomly shuffled data, $\Delta\alpha_\textrm{G}<0.1$ (not shown). (b) Comparison of the multifractal spectra across two regimes: quasiperiodic and hyperchaotic. (c) Corresponding relative frequency histogram of the H\"older exponents.}
\label{Fig3}
\end{figure}

The multifractal analyses for the time-series of $w$ generated as a function of the control parameter $\delta$ are depicted in Fig.~\ref{Fig3}. The parameter was swept in $\delta\in[0.6,1]$, thus allowing the system to develop a wide range of dynamical behaviors comprising both chaotic motions and closed orbits \cite{saito1990}. The corresponding averaged Hurst exponent $H$ and multifractal spectrum width $\Delta \alpha_\textrm{G}$ as estimated through the MFDFA and WL algorithms are depicted in Fig.~\ref{Fig3}a. It is evident that the multifractal characteristics are insensitive to the qualitative features of the system dynamics. The time-series remain strongly persistent, with $H \approx 1.5$, and feature a wide spectrum with $\Delta \alpha_\textrm{G} \approx 2.25$: this could, at the surface, suggest a multifractal organization. In Fig. \ref{Fig3}b, the multifractal spectra for the hyperchaotic and quasiperiodic regimes are compared. Their shape is almost identical, with a strong left-sided asymmetry $A_\alpha \approx 0.5$: importantly, this coexists with an uneven distribution of the points along the spectrum, which concentrate mainly towards its ends. Here, analysis of the local scaling revealed fundamental subtleties of the data organization. The relative frequency histogram $f_r$ of the H\"older exponents $\alpha_\textrm{L}$ forms two separable peaks, whose locations coincide with high-concentration points close to the minimal and maximal values of $\alpha _\textrm{G}$ identified on the multifractal spectrum.

\begin{figure}[h!]
\centering
\includegraphics[scale=0.35,keepaspectratio=true,trim={2cm 0 0 0}]{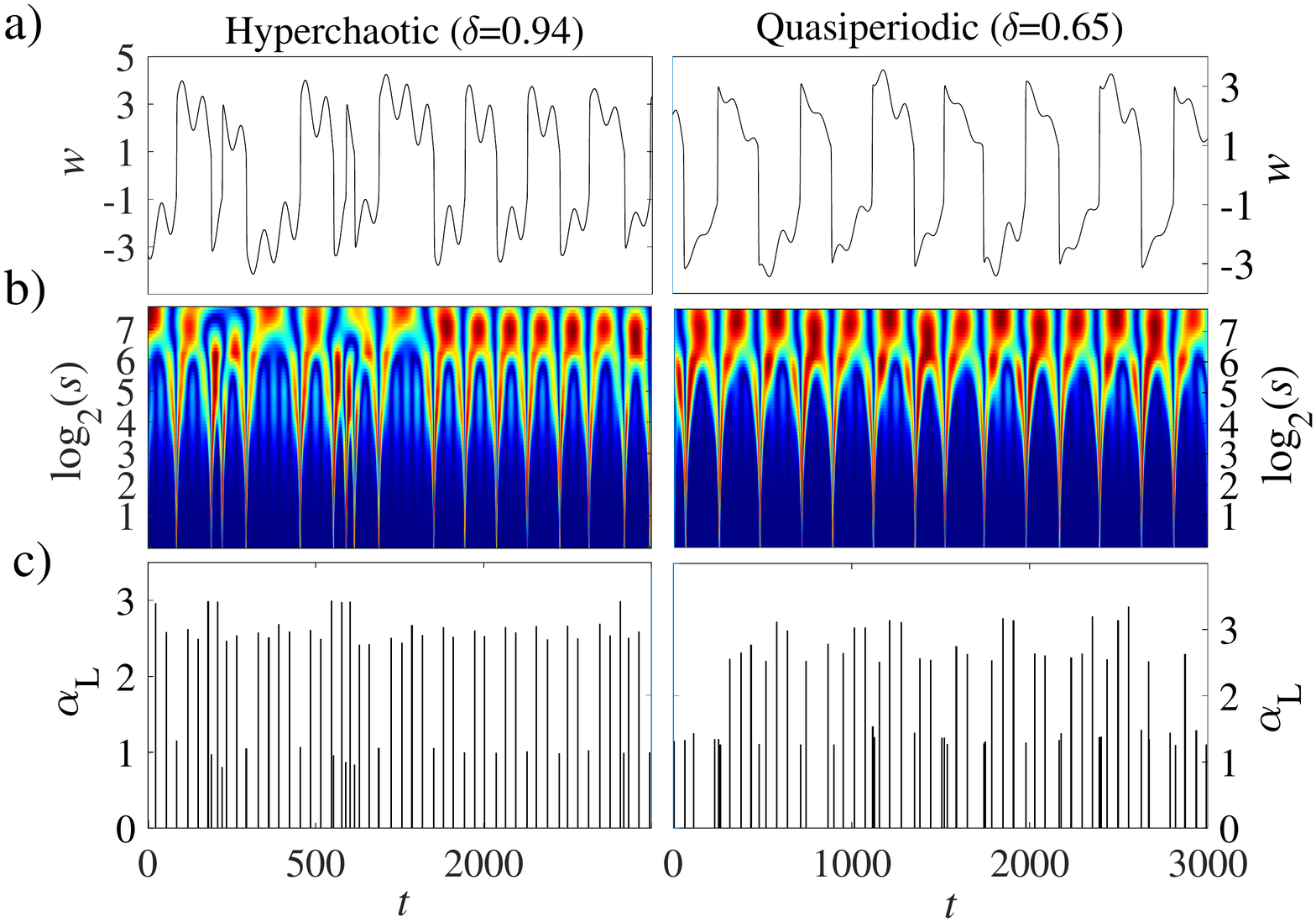}
\caption{Local scaling properties of the $w$ time-series from the Saito chaos generator, simulated in the hyperchaotic (left) and  quasiperiodic (right) regimes. (a) Time-courses and (b) their wavelet transforms obtained via the fourth derivative of the Gaussian wavelet. Color coding denotes the magnitude of the wavelet coefficients $Wf(x_0,s)$ ranging from dark blue (the smallest $Wf(x_0,s)$) to red (the largest one). (c) Corresponding time-localized H\"older exponents.}
\label{Fig4}
\end{figure}

The locations of the singularities and their ``strength'' were recovered, as given by the H\"{o}lder exponents, through analysis of the local wavelet transform coefficients (Fig.~\ref{Fig4}). It is well-evident that for the time-series of the $w$ variable, in both the quasiperiodic and hyperchaotic regimes (Fig.~\ref{Fig4}a), the maxima form separate lines on the space-scale half-plane (cf. Fig.~\ref{Fig4}b), which delineate isolated singularities (Fig.~\ref{Fig4}c). In the presence of a truly multifractal geometry, the maxima would follow a tree-like structure, stemming from the self-similar organization of the fluctuations. By contrast, consideration of the locations and strength of the singularities reveals that two discrete types are present in these time-series, and related to volatile portions of the signal: one reflects instants wherein the hysteretic behavior is apparent ($\alpha_\textrm{L} \approx 1.2$), the other reflects the local extrema of the oscillatory component ($\alpha_\textrm{L} \approx 3$). Thus, a faithful reconstruction of the multifractal spectrum would clustered around two separate points. However, as discussed below, the $q$-filtering method inherently yields a concave spectrum, and isolated peaks are impossible to obtain. The artifactual result, then, is purely the product of the averaging procedures inherent in the MFDFA methodology: together with the dense sampling of the $q$ parameter, these generate a broad spectrum of H\"older exponents, even when only isolated singularities are present in the signal. Although it is markedly stretched towards the left-hand side, with a high concentration of points towards the two limit values of H\"{o}lder exponent, the estimated spectrum resembles an inverted parabola. Instead, the scaling properties of the time-series generated by this system should be represented by a single exponent for the $x,\ y,\ z$ variables, and by a bifractal organization for the $w$ variable.

\subsubsection{Experimental confirmation}
\begin{figure}[h!]
\centering
\includegraphics[scale=0.2,width=0.5\textwidth,keepaspectratio=true]{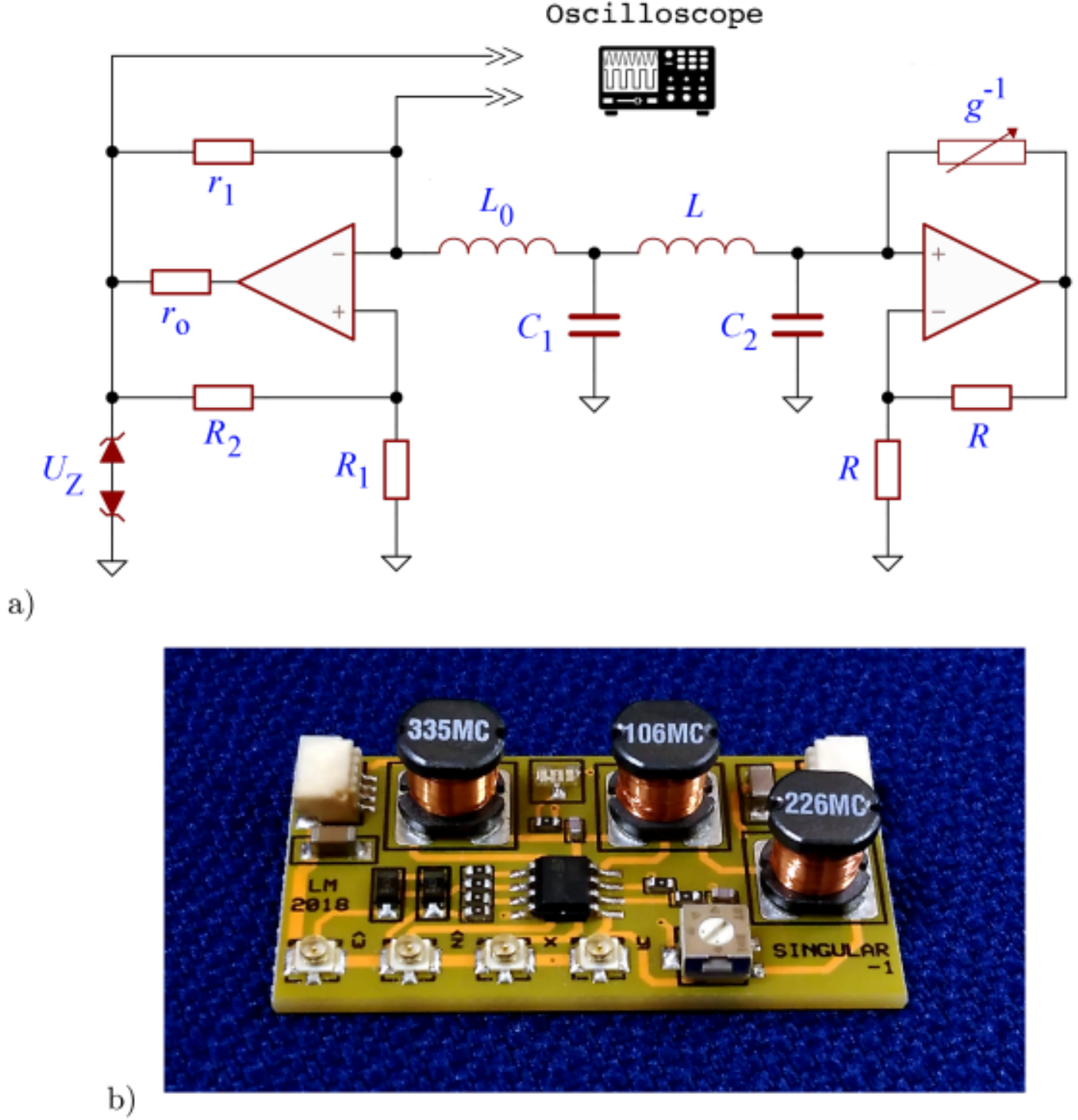}
\caption{Experimental implementation of the Saito chaos generator. (a) Circuit diagram, and (b) representative example of physical realization.}
\label{Fig5}
\end{figure}

To independently confirm that the results presented above stem faithfully from the dynamics of this system, an experimental version of the same was conveniently constructed using two operational amplifiers (type TL082) and a non-linearity based on two anti-parallel series Zener diodes (type BZT52-C5V1). The corresponding circuit diagram is given in Fig.~\ref{Fig5}a, where $r_1=r_2=R_1=R_2=R=10\ \textrm{k}\Omega$, $r_\textrm{o}=820\ \Omega$, $C_1=C_2=3.9\textrm{ nF}$, $L_0=3.3\textrm{ mH}$, $L=32\textrm{ mH}$ (two inductors in series), and $U_\textrm{Z}=5.1\textrm{ V}$. These component values yield $\gamma=C_1/C_2=1$, $\varepsilon=L_0/(r_1^2C_1)=0.0085$, $\eta=r_1/r_2=1$, and $\rho=r_1^2C_1/L=12.2$. The signal corresponding to the variable $w$ was digitized from the physical circuit board (Fig.~\ref{Fig5}b) using a recording oscilloscope at a rate of 1 MSa/s, tuning $g^{-1}$ to obtain $\delta = \{0.63, 0.67, 0.71, 0.77, 0.83, 0.91, 1\}$. The corresponding time-series have been made publicly available \cite{supplmat}.\par

\begin{figure}[h!]
\centering
\includegraphics[scale=0.32,keepaspectratio=true,trim={1.5cm 0 0 0}]{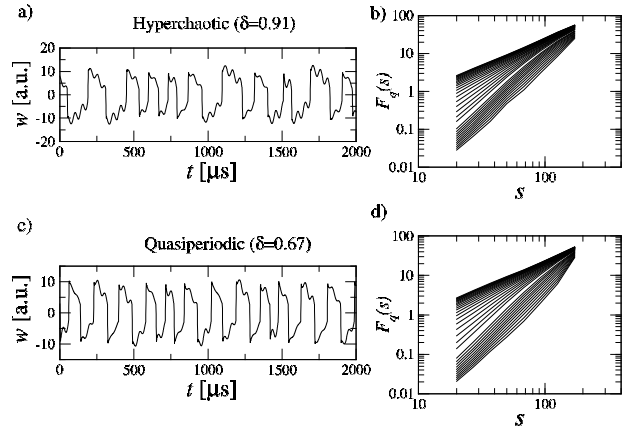}
\caption{(a) Experimental time-series of variable $w$ recorded from the physical Saito chaos generator in the hyperchaotic regime and (b) corresponding fluctuation functions $F_q(s)$. (c) Time-series for the same recorded in the quasiperiodic regime and (d) corresponding fluctuation functions. Y-axis presented in arbitrary units.}
\label{Fig6}
\end{figure}

\begin{figure}[h!]
\centering
\includegraphics[scale=0.35,keepaspectratio=true,trim={1cm 0 0 0}]{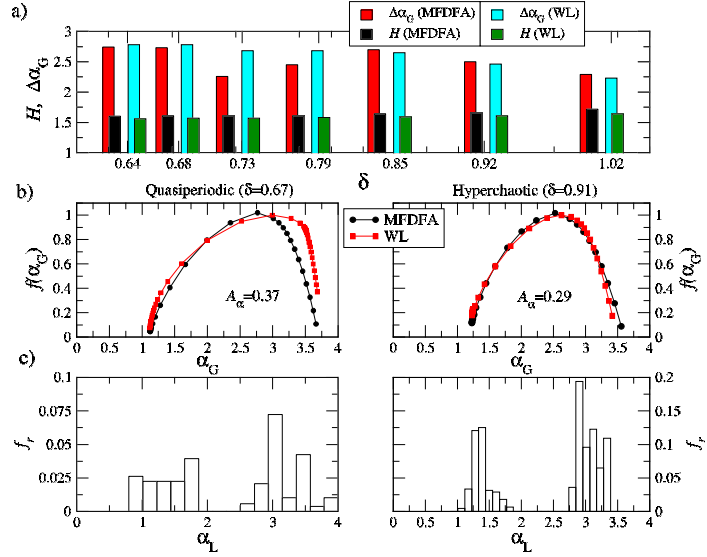}
\caption{Analysis of the experimental recordings from the electronic Saito chaos generator. (a) Width of the multifractal spectrum and Hurst exponents estimated for the $w$ variable given different settings of the control parameter $\delta$. 
Average $\Delta \alpha_\textrm{G}$ for the Fourier based-surrogates and randomly shuffled data, respectively, $<0.65$ and $<0.1$. (b) Multifractal spectra estimated for the quasiperiodic and hyperchaotic regimes. (c) Corresponding relative frequency histograms for the H\"older exponents.}
\label{Fig7}
\end{figure}

In agreement with the simulations, apparent multifractality only arises for the variable $w$. In Fig.~\ref{Fig6}, the time-series in the quasiperiodic and hyperchaos regimes are shown alongside the corresponding fluctuation functions. The heterogeneity of the latter is equally apparent in both cases, suggesting that the multifractal properties of the signals are only weakly dependent on the dynamics. The multifractal spectra are shown in Fig. ~\ref{Fig7}: they are wide ($\Delta \alpha_\textrm{G} > 2$) but, in contrast to the simulations, more symmetric ($A_{\alpha} \approx 0.3$) (cf. Fig.~\ref{Fig7}b). On the other hand, analysis of the local scaling properties reveals a singularity organization comparable to the simulations (cf. Fig.~\ref{Fig7}c). Thus, the variability of the H\"older exponents for the experimental signals is higher; however, the histogram still forms two separable clusters (albeit more dispersed than in the simulations), wherein high values reflect local signal maxima and small ones correspond to the sudden jumps. A possible explanation for the more diverse distribution of the singularities could be sought in the tolerances and non-ideal behaviors of the electronic components (e.g., finite quality factor and self-resonance of the inductors, smooth response of the diodes, etc.), which knowingly give rise to richer dynamics. Altogether, these results confirm the above conclusions, reassuring that they are not an artifact of the numerical integration.

\subsection{Artefactual multifractality in the R\"{o}ssler system}

The next example of dynamical system that we consider is the R\"{o}ssler system. Its dynamics are governed by the following system of three differential equations \cite{rossler1976}

\begin{equation}
\label{Rossler_eqx}
\begin{cases}
 \dot{x}=-y-z\\
 \dot{y}=x+ay\\
 \dot{z}=b+z(x-c)\textrm{  .}\\
 \end{cases}
\end{equation}

\begin{figure}[h!]
\centering
\includegraphics[scale=0.35,keepaspectratio=true,trim={1.5cm 0 0 0}]{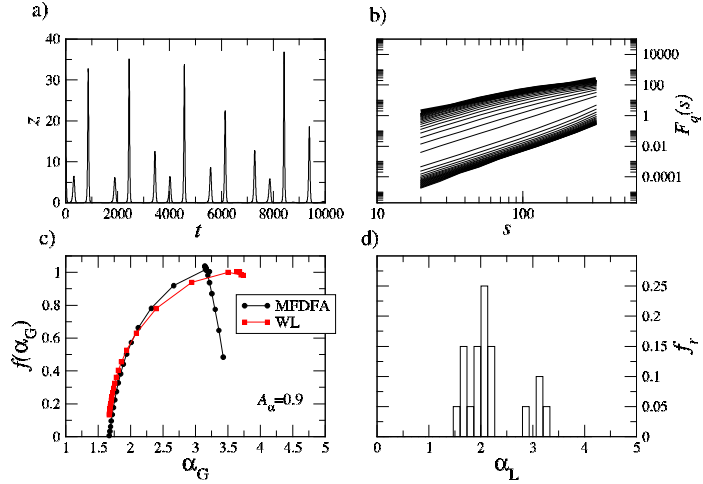}
\caption{Analysis of the simulated $z$ time-series from the R\"{o}ssler system. (a) Representative fragment of the times series. (b) Fluctuation functions $F_q(s)$ from MFDFA algorithm. (c) Multifractal spectra estimated for the analysed signal. (d) Corresponding relative frequency histogram of the H\"older exponents.}
\label{Fig8}
\end{figure}

Similarly to the Saito generator, the R\"{o}ssler system reveals rich dynamics spanning periodic and chaotic behaviors. Notably, its dynamical properties depend on state equations without a hysteresis element. They are controlled by parameters that were set to $a$=0.3, $b$=0.2, $c$=5.7, knowingly realizing chaotic behavior with an intermediate level of folding. To ensure statistical reliability, we generated time-series having a length of $10^6$ points, a representative fragment of which is visible in Fig.~\ref{Fig8}a. The distribution of the fluctuation functions already suggests a bi-fractal organization (Fig.~\ref{Fig8}b) \cite{krawczyk2019}. However, the multifractal spectra estimated through the MFDFA and WL algorithms resemble typical multifractal characteristics, with a strong left-sided asymmetry (Fig.~\ref{Fig8}c). This is in contrast with the local scaling properties (Fig.~\ref{Fig8}d). The histogram of local H\"{o}lder exponents forms, similarly to the Saito generator case, two separate peaks that concentrate in the vicinity of the extreme $\alpha_\textrm{G}$ values. Thus, the underlying structure reflects isolated singularities rather than a unified multifractal organization.  

\subsection{Further examples of artefactual multifractality in synthetic signals}

The results presented above suggest that singular behavior in the Saito chaos generator can be quantified through just two scaling exponents. Moreover, the subsets corresponding to different singularities index separate components of the time-series, rather than constituting hierarchically-interwoven structures, which are the hallmark of true multifractality. Thus, a naive interpretation of the spectrum width $\Delta \alpha_\textrm{G}$ as a signature of multifractality can be faulty. To highlight this issue even more clearly, we finally consider processes that are, by construction, not multifractal. Yet, the methods based on $q$-filtering, namely the MFDFA and WL, yield misleadingly wide multifractal spectra in these cases.\par

\begin{figure}[h!]
\centering
\includegraphics[scale=0.35,keepaspectratio=true,trim={1.5cm 0 0 0}]{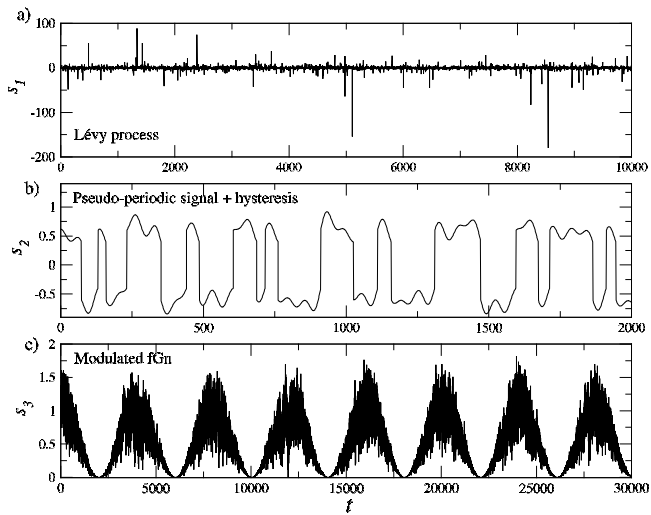}
\caption{Examples of synthetic time-series leading to apparent multifractality. (a) L\'evy process, (b) pseudo-periodic signal with sudden jumps (mimics $w$ variable in the Saito chaos generator), (c) pseudo-multifractal process based on fractional Gaussian noise.}
\label{Fig9}
\end{figure}

\begin{figure}[h!]
\centering
\includegraphics[scale=0.35,keepaspectratio=true,trim={2cm 0 0 0}]{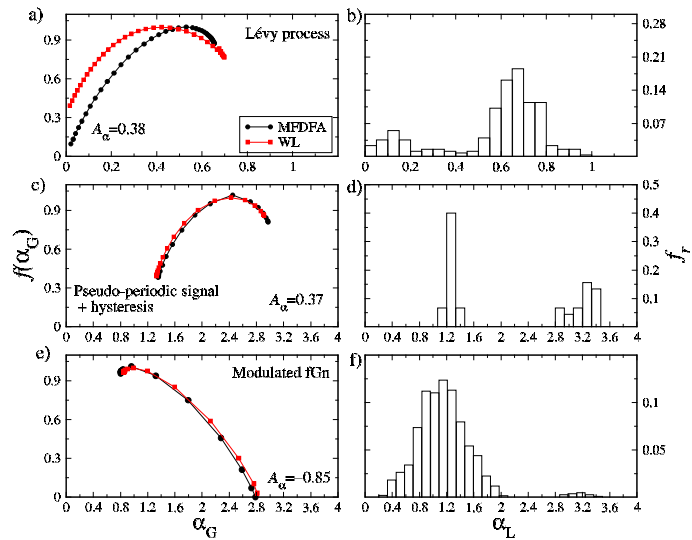}
\caption{Analysis of the synthetic time-series, with multifractal spectra (left) and relative frequency histogram $f_r$ of the H\"older exponents (right). (a) L\'evy process, (b) pseudo-periodic signal with sudden jumps (mimics $w$ variable in the Saito chaos generator), (c) pseudo-multifractal process based on fractional Gaussian noise.}
\label{Fig10}
\end{figure}

As an instructive example, results from the multifractal analysis of the L\'evy process, which possesses a well-recognized bifractal structure, are firstly presented. The multifractal spectrum of the L\'evy time-series consists of two points, whose locations are directly related to the asymptotic behavior of the distribution tail $P(x)\sim x^{-(\alpha_{\textrm{Levy}}+1)}$ and are given by \cite{nakao2000,oswiecimka2006}:
\begin{equation}
\alpha = \left\{ \begin{array}{cc} 1/\alpha_\textrm{Levy} & (q \le \alpha_\textrm{Levy}) \\
0 & (q > \alpha_\textrm{Levy}) \end{array} \right. \ \ 
f(\alpha) = \left\{ \begin{array}{cc} 1 & (q \le \alpha_\textrm{Levy}) \\
0 & (q > \alpha_\textrm{Levy}) \end{array} \right.
\end{equation}
where $\alpha_\textrm{Levy}$ is the L\'evy index and $q$ is $q$-th moment of the fluctuation function $F_q(s)$. The multifractal analysis of the L\'evy time-series having a length of 50000 points (Fig.~\ref{Fig9}a) with $\alpha_\textrm{Levy}=1.5$ is reported in Fig.~\ref{Fig10}a,b. Therein, the bifractal nature of the data is clearly visible in the histogram of H\"older exponents: two peaks, corresponding to the theoretical values, can be readily identified. The dispersion of these peaks is artifact purely due to the finite time-series length. Yet, the multifractal spectrum estimated utilizing the MFDFA and WL methods is wide (i.e., $\Delta \alpha_\textrm{G} = 0.7$) and strongly left-sided asymmetrical ($A_\alpha \approx 0.38$): this could lead to the faulty conclusion that the process is multifractal when, in reality, it is not.\par

Next, a minimal-complexity arrangement which can reproduce qualitative characteristics similar to those observed in the Saito chaos generator is considered: it simply consists of the linear superposition $w(t)$ of two related signals. One is a  pseudo-periodic signal given by, e.g., $u(t)=\sum_i\sin2\omega t/(p_i/\max p)$ where $p_i=\{2,3,5,7,11\}$. The other is a sequence of binary fluctuations $v(t)=\mathcal{W}[u(t),\xi]$ generated by a hysteresis operator $\mathcal{W}$ acting on that signal, with $v\in[-1,1]$ and hysteresis parameter $\xi=0.2\max u$. Their linear combination, e.g., $w(t)=v(t)+u(t)/\max u$ is, by definition, not hierarchically interwoven and does not obey different scaling exponents \cite{hu2001,ludescher2011}, hence, the multifractal spectrum should consist of two separable points. To test this hypothesis, 10 time-series segments each having a lengths of $10^6$ points were generated, and MFDFA was performed. The average of the multifractal spectra and the histogram of the H\"older exponents are depicted in Fig. \ref{Fig10}c,d. In this case too, the multifractal spectrum appears well-developed ($\Delta \alpha_\textrm{G} = 1.5$), with a strong left-sided asymmetry ($A_\alpha = 0.37$), hence cursory interpretation of these results might suggest a complex multifractal structure. Again, the true nature of the process is revealed by the histogram of the H\"older exponents estimated through the wavelets, which shows that only two discrete types of singularities are present in the time-series. Thus, the analyzed structure is closer to a fractal structure than to multifractality. It is worth noting that the distribution of the H\"older exponents as well as the shape of the multifractal spectrum resemble closely the results for the variable $w$ in the Saito chaos generator; here, however, there was no underlying non-linear dynamical system.
 
Finally, a pseudo-multifractal process, consisting of the superposition of a fractal time-series with periodic components, is considered. At a first glance, this process resembles the multifractal time-series of sunspot number variability (cf. Fig.\ref{Fig1}). However, as demonstrated below, careful inspection of the fractal characteristics illuminates its pseudo-multifractality. To this end, fractional Gaussian noise (fGn) \cite{mandelbrot1968} was generated: it represents a well-known example of a stochastic monofractal structure with possible long-range correlations quantified by the Hurst exponent, which has been applied to model phenomena across various fields of science. Namely, simulations produced a fractional Gaussian noise with an arbitrarily-chosen Hurst exponent of $H=0.8$ (strongly persistent behavior), for which the amplitude of the process was modulated by a cosinusoidal function $F(i)=A+A\cos(2\pi i/T_0)$ in $i=1...N$, where $A$ and $T_0$ are the model parameters (cf. Fig.~\ref{Fig9}c). Then, the periodic function $F(i)$ was added to this amplitude modulated noise. In our simulations, $N=10^6$, $A=0.5$ and $T_0=4000$ were set. The results of the local scaling analysis, as well as the multifractal spectrum, are shown in Fig. \ref{Fig10}e,f. Analysis of the distribution of H\"older exponents confirms that the time-series is a composition of the two independent processes having different singular behaviors. The smaller values of $\alpha_\textrm{L}$ concentrate around $\alpha_\textrm{L} \approx H = 0.8$, corresponding to the fGn component, whereas the larger ones are related to the periodic trend. A cursory analysis of the multifractal spectrum indicates heterogeneity in the scaling properties. Particularly, the width of the spectrum, together with its strong right-sided asymmetry ($A_\alpha=-0.85$), could be taken as the hallmark of a multifractal time-series with a well-developed hierarchy of small fluctuations; however, this is purely an artifact. To avoid mistaking pseudo-multifractality as a valid multifractal structure, the analysis of the global fractal properties should always be corroborated by consideration of the local scaling properties as given by the wavelet transform.

\begin{figure}[h!]
\centering
\includegraphics[scale=0.35,keepaspectratio=true,trim={2cm 0 0 0}]{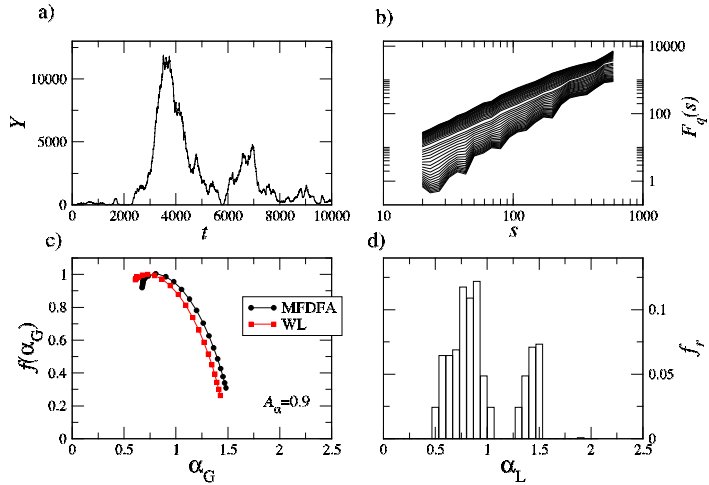}
\caption{Analysis of the square transform of fractional Brownian motion. (a) Representative fragment of the times series. (b) Fluctuation functions $F_q(s)$ from MFDFA algorithm. (c) Multifractal spectra estimated for the analysed signal. (d) Corresponding relative frequency histogram of the H\"older exponents.}
\label{Fig11}
\end{figure}

As a last example, we analyzed an artefactually-generated stochastic process with the singularity spectrum derived analytically. In this respect, we considered the square transform of fractional Brownian motion (Fig.~\ref{Fig11}a), which represents a bi-Hölder process whose spectrum is given by the following relation \cite{abry2015}:
\begin{equation}
f(\alpha) = \left\{ \begin{array}{lcl} 1 & \textrm{if} & \alpha = H \\
1-H & \textrm{if} & \alpha=2H \\ 
-\infty & & \textrm{elsewhere}. \end{array} \right.
\end{equation}
In our study, we generated fBm having a length of $10^6$ points with a Hurst exponent $H=0.7$. Fluctuations functions $F_q(s)$ (Fig.~\ref{Fig11}b) obtained through MFDFA show non-homogeneous scaling, which suggests a multiscaling behavior of the data. This is even more clearly visible in the singularity spectrum, which reproduces a concave hull supported by the interval in the range from $H$ to $2H$ (Fig.~\ref{Fig11}c). Thus, based only on the Legendre-based methodology, a flawed conclusion on the multifractal structure of the data would be drawn. However, consideration of the histogram of H\"older exponents (Fig.~\ref{Fig11}d) estimated through wavelet analysis reveals the true bi-fractal nature of the process, with exponents corresponding to the theoretical expectations. This leads to the conclusion that the exponents identified through MFDFA and WL, except the two extreme values, are an artifact caused by a methodological limitation and do not contain any true information about the analysed process.

\section{Discussion}

The present study was initially conceived as the search for a deterministic non-linear dynamical system, relatively low-dimensional, which could genuinely generate multifractal structures. The Saito chaos generator, representing a hysteretic oscillator which can be easily simulated numerically as well as realized experimentally in an electronic circuit, was identified as a promising candidate. Indeed, an initial investigation of its dynamics based on the commonly-accepted MFDFA technique yielded a spectrum suggestive of fully-developed multifractality; similar conclusions were drawn from the WL analysis. However, concerningly, these results could only be obtained for the dynamical variable $w$, which features discontinuities in the form of jumps, and not for the others. A clear indication of a flaw came from the observation of largely overlapping multifractal spectra between the hyperchaotic and quasiperiodic regimes: since in the latter there is no turbulence, this is unexpected.\par

Closer visual inspection of the time-series was instrumental in resolving this issue, because it demonstrated the absence of a cascaded hierarchy of singularities, fractally nested over the consecutive scales of magnification; instead, it revealed merely a sequence of isolated singularities, associated with a restricted number of distinct H\"older exponents. In the Saito chaos generator, these isolated singularities are produced by twofold dynamics of the system, namely a continuous manifold and sudden jumps. The dynamics of these components, whose organization is quantified by single scaling exponents, are not strongly interrelated: this results in a fluctuation structure devoid of hierarchical organization. An explicit step-by-step wavelet-based estimation of the same indicated values limited to two-well separated, narrow intervals (Fig.~\ref{Fig3}): this is in stark contrast with the results given by both the MFDFA and WL algorithms, which generated a broad distribution of $f(\alpha)$, comprising the two H\"older exponents seen locally, but mistakenly suggesting coverage of the entire interval between these extremes. Tentatively, a combination of processes without intrinsic convolution can be indicated as a necessary condition for the emergence of artefactual multifractality related to isolated singularities. Thus, our methodology can be applied to dynamical systems revealing different forms of dynamics (cf. Fig.~\ref{Fig7}); for instance, the Saito system in the (quasi-)periodic regime is characterized by two types of behavior. One, related to the periodic component, and the other, produced by hysteresis. However, these dynamics are not hierarchically nested, which leads to a set of isolated singularities, which in turn masquerade as a uniform multifractal structure. Further synthetic signals revealed that the danger of over-evaluating the multifractal content is not confined to the dynamics of this particular oscillator: quite on the contrary, the diversity of these signals points to a potentially pervasive problem.\par

At the same time, for the genuine multifractals considered, such as the binomial cascade, the MFDFA and WL methods reproduced remarkably closely the explicitly-determined distribution of underlying H\"older exponents. Similarly, a good correspondence was found for the other cases of fully-developed multifractality, such as the series of heart inter-beat intervals or sentence length variability in ``Finnegans Wake''. These methods, then, are obviously not per-se flawed, but their results need to be interpreted much more cautiously than is generally done.\par

However, precise numerical determination of the singularity spectrum for experimental signals knowingly remains an open problem \cite{cardo2009}. Due to the intricate nature of the multifractal formalism, there are no theoretical proofs of the mathematical validity of the algorithms used to estimate the singularity spectrum: these algorithms are treated simply as reasonable numerical approximations of the underlying ground truth \cite{kantelhardt2012}. In particular, the commonly used formulas for the multifractal spectrum have been proposed heuristically, through analogy with thermodynamic formalism, and verified numerically for specific cases based on multifractal measures \cite{jaffard1997b}. Generally, the multifractal formalism yields the upper bound of the singularity spectrum as was shown in \cite{jaffard1997a}, but not the exact spectrum. Therefore, it does not lend itself to formal proof. The present demonstration of what cases under which the MFDFA and WL algorithms produce a flawed indication of multifractality provides essential inspiration and input for further, more mathematically rigorous inquiry into the validity of these methods.\par

It is important to point out that the observed disagreement between the multifractal spectra and the true distribution of the underlying H\"older exponents is not merely down to avoidable algorithmic implementation choices, but stems straight from the foundational assumptions upon which the multifractal analyses under consideration are built. The MFDFA and WL approaches represent two related advances of the Parisi-Frisch procedure \cite{parisi1985}, aiming for a compromise between reflecting the mathematical definition of the H\"older exponent and obtaining an estimate which is practically viable in terms of both computational load and stability. An averaging procedure related to the partition function (or its equivalent, cf. Eqs.~(\ref{Fluctuations_funtion}) and (\ref{partition_function})) and Legendre transform (cf. Eq.~(\ref{spectrum-LW})) is central to their framework. Although numerically stable, this methodology has one serious drawback. Namely, the Legendre transform by construction imposes the concavity of the multifractal spectrum \cite{cardo2009} a priori even when this is not true: if non-concave H\"older characteristics are considered, it provides only the upper bound of the H\"older spectrum \cite{leonarduzzi2016}. Therefore, it is patently impossible to say without additional tests whether an observed concave spectrum is a valid representation of the data, or simply reflects the limitations of the methodology. To overcome this problem, multifractal methods not based on the Legendre transform have to be applied \cite{esser2017,turiel2006}. However, in many cases, these methods suffer from issues of practical implementation and computational stability. Therefore, in this work, the distribution of the H\"older exponents was instead directly analyzed. In the presence of non-concave multifractal characteristics, it formed two clearly-separable clusters: this inexorably illuminates the true organization of the data and thus demonstrates the risk of faulty interpretation of MFDFA- and WL-based results.\par

To conclude, while these techniques offer very efficient and practical tools for quantifying the principal characteristics of multifractal patterns both in time and in space, extreme care needs to be taken when a suspicion arises that the pattern does not stem from true, uniform multifractality. The reason is that the methods are, by design, forced to assume a multifractal form for the singularity spectrum. This aspect points to a definite necessity to be addressed in future developments, namely, mitigating this strong prior assumption. For now, inspection of the distribution of the local H\"older exponents as given by wavelet-based techniques is crucial in identifying such instances and classifying them correctly.

\section*{Acknowledgement}
L. Minati gratefully acknowledges funding by the World Research Hub Initiative (WRHI), Institute of Innovative Research (IIR), Tokyo Institute of Technology, Tokyo, Japan. L. Zunino was supported by the Argentinean Consejo Nacional de Investigaciones Cient\'ificas y T\'ecnicas (CONICET).


\begin{thebibliography}{999}

\bibitem{pietronero2012}Pietronero, L., Tosatti, E.:
Fractals in Physics.
North Holland (2012)  

\bibitem{kwapien2012}Kwapie\'n, J., Dro\.zd\.z, S.:
Physical approach to complex systems.
Phys. Rep. \textbf{515}, 115--226 (2012)

\bibitem{vicsek1992}Vicsek, T.:
Fractal Growth Phenomena.
World Scientific (1992)

\bibitem{mandelbrot1968}Mandelbrot, B.B., Van Ness, J.W.:
Fractional Brownian motions, fractional noises and applications.
SIAM Rev. \textbf{10}, 422--437 (1968)

\bibitem{peng1992}Peng, C.-K., Buldyrev, S.V., Goldberger, A.L., Havlin, S., Sciortino, F., Simons, M., Stanley, H.E.:
Long-range correlations in nucleotide sequences.
Nature \textbf{356}, 168--170 (1992) 

\bibitem{frisch1985}Frisch, U., Parisi, G.:
Fully developed turbulence and intermittency.
In:  Ghil, Michael (ed.) Turbulence and Predictability in Geophysical Fluid Dynamics and Climate Dynamics. North-Holland, New York (1985)

\bibitem{halsey1986}Halsey, T.C., Jensen, M.H., Kadanoff, L.P., Procaccia, I., Shraiman, B.I.:
Fractal measures and their singularities: The characterization of strange sets.
Phys. Rev. A \textbf{33}, 1141--1151 (1986)

\bibitem{drozdz2018}Dro\.zd\.z, S., Kowalski, R., O\'swi\c ecimka, P., Rak, R., G\c ebarowski, R.: 
Dynamical variety of shapes in financial multifractality. 
Complexity \textbf{2018}, article ID 7015721 (2018)

\bibitem{muzy2008}Muzy, J.F., Bacry, E., Baile, R., Poggi, P.:
Uncovering latent singularities from multifractal scaling laws in mixed asymptotic regime. Application to turbulence.
EPL(Europhysics Letters) \textbf{82}, 60007 (2008)

\bibitem{subramaniam2008}Subramaniam, A.R., Gruzberg, I.A., Ludwig, A.W.W.:
Boundary criticality and multifractality at the two-dimensional spin quantum Hall transition.
Phys. Rev. B \textbf{78}, 245105 (2008)

\bibitem{ivanov1999}Ivanov, P.Ch., Amaral, L.A.N., Goldberger, A.L., Havlin, S., Rosenblum, M.G., Struzik, Z.R., Stanley, H.E.:
Multifractality in human heartbeat dynamics. 
Nature \textbf{399}, 461--465 (1999)

\bibitem{rosas2002}Rosas, A., Nogueira Jr., E., Fontanari, J.F.:
Multifractal analysis of DNA walks and trails.
Phys. Rev. E \textbf{66}, 061906 (2002)

\bibitem{makowiec2009}Makowiec, D., Dudkowska, A., Ga{\l}\c{a}ska, R., Rynkiewicz, A.:
Multifractal estimates of monofractality in RR-heart series in power spectrum ranges.
Physica A \textbf{388}, 3486--3502 (2009)

\bibitem{stanley1999}Stanley, H.E., Amaral, L.A.N., Goldberger, A.L., Havlin, S., Ivanov,  P.Ch., Peng, C.-K.:
Statistical physics and physiology: Monofractal and multifractal approaches.
Physica A \textbf{270}, 309--324 (1999)

\bibitem{franca2018}Franca, L.G.S., Miranda, J.G.V., Leite, M., Sharma, N.K., Walker, M.C., Lemieux, L., Wang, Y.:
Fractal and Multifractal Properties of Electrographic Recordings of Human Brain Activity: Toward Its Use as a Signal Feature for Machine Learning in Clinical Applications.
Front. Physiol. \textbf{9}, 1767 (2018)

\bibitem{stanley1988}Stanley, H.E., Meakin, P.:
Multifractal phenomena in physics and chemistry.
Nature \textbf{335}, 405--409 (1988)

\bibitem{udovichenko2002}Udovichenko, V.V., Strizhak, P.E.:
Multifractal Properties of Copper Sulfide Film Formed in Self-Organizing Chemical System.
Theor. Exp. Chem. \textbf{38}, 259--262 (2002)

\bibitem{ausloos2002}Ausloos, M., Ivanova, K.:
Multifractal nature of stock exchange prices.
Comput. Phys. Commun. \textbf{147}, 582--585 (2002)

\bibitem{turiel2005}Turiel, A., Perez-Vicente, C.J.:
Role of multifractal sources in the analysis of stock market time-series.
Physica A \textbf{355}, 475--496 (2005)

\bibitem{kutner2019}Kutner, R., Ausloos, M., Grech, D., Di Matteo, T., Schinckus, C., Stanley, H.E.:
Econophysics and sociophysics: Their milestones \& challenges.
Physica A \textbf{516}, 240--253 (2019)

\bibitem{oswiecimka2005}O\'swi\c ecimka, P., Kwapie\'n, J., Dro\.zd\.z, S.:
Multifractality in the stock market: price increments versus waiting times.
Physica A \textbf{347}, 626--638 (2005)

\bibitem{ruan2011}Ruan, Y.-P., Zhou, W.-X.:
Long-term correlations and multifractal nature in the intertrade durations of a liquid Chinese stock and its warrant.
Physica A \textbf{390}, 1646--1654 (2011)

\bibitem{grech2013}Grech, D., Pamu{\l}a, G.:
Multifractality of Nonlinear Transformations with Application in Finances.
Acta Phys. Pol. A \textbf{123}, 529--537 (2013)

\bibitem{gebarowski2019}G\c ebarowski, R., O\'swi\c ecimka, P., W\c{a}torek, M., Dro\.zd\.z, S.:
Detecting correlations and triangular arbitrage opportunities in the Forex by means of multifractal detrended cross-correlations analysis.
Nonlinear Dyn. \textbf{98}, 2349--2364 (2019) 

\bibitem{ausloos2012}Ausloos, M.:
Generalized Hurst exponent and multifractal function of original and translated texts mapped into frequency and length time-series.
Phys. Rev. E \textbf{86}, 031108 (2012) 

\bibitem{drozdz2016}Dro\.zd\.z, S., O\'swi\c ecimka, P., Kulig, A., Kwapie\'n, J., Bazarnik, K., Grabska-Gradzi\'nka, I., Rybicki, J.,  Stanuszek, M.:
Quantifying origin and character of long-range correlations in narrative texts.
Inform. Sci. \textbf{331}, 32--44 (2016)

\bibitem{jafari2007}Jafari, G.R., Pedram, P., Hedayatifar, L.:
Long-range correlation and multifractality in Bach's Inventions pitches. 
J. Stat. Mech: Theory Exp. \textbf{2007}, P04012 (2007)

\bibitem{oswiecimka2011}O\'swi\c ecimka, P., Kwapie\'n, J., Celi\'nska, I., Dro\.zd\.z, S., Rak, R.:
Computational approach to multifractal music. 
arXiv:1106.2902 (2011)

\bibitem{jiang2019}Jiang, Z.-Q., Xie, W.-J., Zhou, W.-X., Sornette, D.:
Multifractal analysis of financial markets: a review.
Rep. Prog. Phys., \textbf{82}, 125901 (2019) 

\bibitem{xiong2017}Xiong, H., Shang, P.: 
Weighted multifractal analysis of financial time series. 
Nonlinear Dyn., \textbf{87}, 2251–2266 (2017)

\bibitem{gu2010}Gu, G.-F., Zhou, W.-X.: 
Detrending moving average algorithm for multifractals. 
Phys. Rev. E \textbf{82}, 011136 (2010) 

\bibitem{kantelhardt2002}Kantelhardt, J.W., Zschiegner, S.A., Koscielny-Bunde, E., Havlin, S., Bunde, A., Stanley, H.E.: 
Multifractal detrended fluctuation analysis of nonstationary time-series. 
Physica A \textbf{316}, 87 (2002)

\bibitem{arneodo1995b}Arneodo, A., Bacry, E., Muzy, J.F.:
The thermodynamics of fractals revisited with wavelets.
Physica A \textbf{213}, 232--275 (1995)

\bibitem{jaffard1998}Jaffard, S.:
Oscillation spaces: Properties and applications to fractal and multifractal functions. 
J. Math. Phys. \textbf{39}, 4129--4141 (1998)

\bibitem{oswiecimka2006}O\'swi\c ecimka, P., Kwapie\'n, J., Dro\.zd\.z, S.: 
Wavelet versus detrended fluctuation analysis of multifractal structures. 
Phys. Rev. E \textbf{74}, 016103 (2006)

\bibitem{salat2017}Salat, H., Murcio, R., Arcaute, E.:  
Multifractal methodology. 
Physica A \textbf{473}, 467-487 (2017)

\bibitem{turiel2006}Turiel, A., P\'{e}rez-Vicente, C.J., Grazzini, J.: 
Numerical methods for the estimation of multifractal singularity spectra on sampled data: A comparative study. 
{J. Comput. Phys.} \textbf{216}, 362--390 (2006)


\bibitem{jaffard2006a}Jaffard, S., Lashermes, B., Abry, P.:
Wavelet Analysis and Applications.
In: Benedetto, J., Frazier, M. (eds.) Applied and Numerical Harmonic Analysis. Springer (2006)

\bibitem{wang2014}Wang, F., Li, Z.-S., Liao, G.-P.:
Multifractal detrended fluctuation analysis for image texture feature representation.
Int. J. Pattern Recognit. Artif. Intell. \textbf{28}, 1455005 (2014) 
 
\bibitem{oswiecimka2018}O\'swi\c ecimka, P., Livi, L., Dro\.zd\.z, S.:
Right-side-stretched multifractal spectra indicate small-worldness in networks.
Commun. Nonlinear Sci. Numer. Simul. \textbf{57}, 231--245 (2018)
  
\bibitem{biswas2012}Biswas, A., Zeleke, T.B., Si, B.C.:
Multifractal detrended fluctuation analysis in examining scaling properties of the spatial patterns of soil water storage.
Nonlinear Proc. Geoph. \textbf{19}, 227--238 (2012)

\bibitem{dutta2013}Dutta, S., Ghosh, D., Chatterjee, S.:
Multifractal detrended fluctuation analysis of human gait diseases.
Front. Physiol. \textbf{4}, 274 (2013)

\bibitem{thompson2016}Thompson, J.R., Wilson, J.R.:
Multifractal detrended fluctuation analysis: Practical applications to financial time-series.
Math. Comput. Simul. \textbf{126}, 63--88 (2016)

\bibitem{kwapien2005}Kwapie\'n, J., O\'swi\c ecimka, P., Dro\.zd\.z, S.:
Components of multifractality in high-frequency stock returns.
Physica A \textbf{350}, 466--474 (2005)

\bibitem{oswiecimka2013}O\'swi\c ecimka, P., Dro\.zd\.z, S., Kwapie\'n, J., G\'orski, A.Z.:
Effect of detrending on multifractal characteristics.
Acta Phys. Pol. A \textbf{123}, 597--603 (2013)

\bibitem{hurst1951}Hurst, H.E.:
Long-term storage capacity of reservoirs.
Trans. Amer. Soc. Civ. Engrs \textbf{116}, 770 (1951)

\bibitem{feder1988}Feder, J.: 
Fractals.
Plenum Press (1998)

\bibitem{calvet1997}Calvet, L., Fisher, A., Mandelbrot, B.B.:
Large Deviations and the Distribution of Price Changes.
In: Cowles Foundation Discussion Paper 1165, Cowles Foundation for Research in Economics. Yale University (1997)

\bibitem{drozdz2015}Dro\.zd\.z, S., O\'swi\c ecimka, P.:
Detecting and interpreting distortions in hierarchical organization of complex time-series.
Phys. Rev. E \textbf{91}, 030902(R) (2015)


\bibitem{muzy1994}Muzy, J.F., Bacry, E., Arneodo, A.:
The multifractal formalism revisited with wavelets.
Int. J. Bifurc. Chaos \textbf{4}, 245--302 (1994)

\bibitem{jaffard2004}Jaffard, S.:
Wavelet techniques in multifractal analysis.
In: Lapidus M., van Frankenhuijsen M. (eds.) Fractal Geometry and Applications: A Jubilee of Benoit Mandelbrot,  Proceedings of Symposia in Pure Mathematics, vol. 72(2),pp. 91--152, AMS (2004)

\bibitem{arneodo1995a}Arneodo, A., Bacry, E., Muzy, J.F.:
Oscillating singularities in locally self-similar functions.
Phys. Rev. Lett. \textbf{74}, 4823--4826 (1995)

\bibitem{lashermes2005}Lashermes, B., Jaffard, S., Abry, P.:
Wavelet leader based multifractal analysis.
Proc. of the Int. Conf. on Acoust. Speech and Sig. Proc. \textbf{4}, 161--164 (2005)

\bibitem{wendt2009}Wendt, H., Roux, S., Jaffard, S., Abry, P.:
Wavelet leaders and bootstrap for multifractal analysis of images.
Signal Process. \textbf{89}, 1100--1114 (2009) 

\bibitem{deliege2014}Deli\`ege, A., Nicolay, S.:
A wavelet leaders-based climate classification of European surface air temperature signals. 
Proceedings of the International work-conference on time-series \textbf{1}, 40--51 (2014)

\bibitem{daubechies1988}Daubechies, I.: 
Orthonormal bases of compactly supported wavelets.
Comm. Pure and Appl. Math. \textbf{41}, 909--996 (1988).

\bibitem{figliola2012}Figliola, A., Rosenblatt, M., Serrano, E.P.: 
Local regularity analysis of market index for the 2008 economical crisis.
Rev. Mat \textbf{19}, 65--78 (2012)

\bibitem{jaffard2006}Jaffard, S., Lashermes, B., Abry, P.: 
Wavelet Leaders in Multifractal Analysis. 
In: Qian, T., Vai, M.I., Yuesheng, X. (eds.) Wavelet Analysis and Applications. Birkh\"{a}user Verlag, Basel (2006)

\bibitem{ayache2004}Ayache, A., L\'{e}vy-V\'{e}hel, J.:
On the identification of the pointwise H\"older exponent of the generalized multifractional Brownian motion.
Stochastic Process. Appl. \textbf{111}, 119--156 (2004) 

\bibitem{schreiber2000}Schreiber, T., Schmitz, A.:
Surrogate time-series.
Physica D \textbf{142}, 346--382 (2000)


\bibitem{drozdz2009}Dro\.zd\.z, S., Kwapie\'n, J., O\'swi\c ecimka, P., Rak, R.: 
Quantitative features of multifractal subtleties in time series. 
EPL (Europhysics Letters) \textbf{88}, 60003 (2009) 

\bibitem{stephen2011}Stephen, D.G., Dixon, J.A.: 
Strong anticipation: Multifractal cascade dynamics modulate scaling in synchronization behaviors. 
Chaos, Solitons \& Fractals \textbf{44}, 160--168 (2011)

\bibitem{saito1990}Saito, T.: 
An approach toward higher dimensional hysteresis chaos generator.
IEEE Trans. Circ. Syst. \textbf{37}, 399--409 (1990)

\bibitem{buscarino2017}Buscarino, A., Fortuna, L., Frasca, M.:
Essentials of Nonlinear Circuit Dynamics with MATLAB$\textregistered$ and Laboratory Experiments. 
CRC Press (2017)

\bibitem{dormand1980}Dormand, J.R., Prince, P.J.:
A family of embedded Runge-Kutta formulae.
J. Comput. Appl. Math. \textbf{6}, 19--26 (1980)

\bibitem{supplmat}Publicly Available Data. Available: http://www.lminati.it/listing/2019/e/,
\textit{Accessed on Oct. 9, 2019}


\bibitem{rossler1976}R\"{o}ssler, O. E.: 
An equation for continuous chaos. 
Phys. Lett. A \textbf{57},  397–-398 (1976)

\bibitem{krawczyk2019}Krawczyk, M.J., O\'swi\c ecimka, P., Kułakowski, K., Dro\.zd\.z, S.: 
Ordered Avalanches on the Bethe Lattice. 
Entropy \textbf{21}, 968 (2019)


\bibitem{nakao2000}Nakao, H.:
Multi-scaling properties of truncated L\'evy flights.
Phys. Lett. A \textbf{266}, 282--289 (2000)

\bibitem{hu2001}Hu, K., Ivanov, P.C., Chen, Z., Carpena, P., Stanley,  H.E.: 
Effect of trends on detrended fluctuation analysis, Phys. Rev. E \textbf{64}, 011114 (2001)

\bibitem{ludescher2011}Ludescher, J., Bogachev, M.I., Kantelhardt, J. W., Schumann, A.Y., Bunde, A.:
On spurious and corrupted multifractality: The effects of additive noise, short-term memory and periodic trends.
Physica A \textbf{390}, 2480--2490 (2011)

\bibitem{abry2015}Abry, P., Jaffard, S., Wendt, H.: A Bridge Between Geometric Measure Theory and Signal Processing: Multifractal Analysis. In: Operator-Related Function Theory and Time-Frequency Analysis. The Abel Symposium 2012. Gröchenig, K., Lyubarskii, Y., Seip, K. (Eds.) Springer, Cham, (2015)

\bibitem{cardo2009}Cardo, R., Corval\'{a}n, A.:
Non-concave multifractal spectra with wavelet leaders projection of signals with and without chirps.
{Fractals} \textbf{17}, 311--322 (2009)

\bibitem{kantelhardt2012}Kantelhardt J.W.: 
Fractal and Multifractal Time Series. 
In: Meyers, R. (Eds.) Mathematics of Complexity and Dynamical Systems. Springer, New York, NY (2012)

\bibitem{jaffard1997b}Jaffard, S.: 
Multifractal formalism for functions part II: Self-Similar Functions. 
SIAM J. Math. Anal. \textbf{28}, 971–998 (1997)

\bibitem{jaffard1997a}Jaffard, S.: 
Multifractal formalism for functions part I: Results Valid For All Functions. 
SIAM J. Math Anal. \textbf{28}, 944–970 (1997)

\bibitem{parisi1985}Parisi, G., Frisch, U.:
On the singularity structure of fully developed turbulence.
In: Turbulence and Predictability in Geophysical Fluid Dynamics. Ghil, M., Benzi, R., G. Parisi, G. (Eds.). pp. 84–87.
Proceedings of the International School of Physics ``Enrico Fermi'' 1983, North-Holland (1985)

\bibitem{leonarduzzi2016}Leonarduzzi, R., Touchette, H., Wendt, H., Abry, P., Jaffard, S.: 
Generalized Legendre transform multifractal formalism for nonconcave spectrum estimation. 
2016 IEEE Statistical Signal Processing Workshop (SSP), 1--5 (2016) 

\bibitem{esser2017}Esser, C., Kleyntssens, T., Nicolay, S.:
A multifractal formalism for non-concave and non-increasing spectra: The leaders profile method.
Appl. Comput. Harmon. Anal. \textbf{43}, 269--291 (2017)

\end{thebibliography}
\end{document}